\documentstyle[12pt]{article}
\begin{document}

\topmargin 0pt

\oddsidemargin -3.5mm

\headheight 0pt

\topskip 0mm
\addtolength{\baselineskip}{0.20\baselineskip}
\begin{flushright}
MIT-CTP-$2728$
\end{flushright}
\begin{flushright}
hep-th/$9805033$
\end{flushright}
\vspace{0.5cm}
\begin{center}
    {\large \bf  (2+1)-Dimensional QED, Anomalous Surface-Term 
Contributions and Superconductivity}
\end{center}
\vspace{0.5cm}
\begin{center}
 Mu-In Park\footnote{Electronic address: mipark@physics.sogang.ac.kr {\it and} mipark@ctpa03.mit.edu} \\
{ Center for Theoretical Physics, Massachusetts Institute of Technology, \\
 Cambridge, Massachusetts 02139 U.S.A.}
\end{center}
\vspace{0.5cm}
\begin{center}
    {\bf ABSTRACT}
\end{center}

It is shown that (2+1)-dimensional QED reveals several unusual effects due to 
the surface-term contributions. It is also shown that this system provides a 
new pairing mechanism for the high-$T_c$ superconductivity on the plane.
\vspace{3cm}
\vspace{2.5cm}
\begin{flushleft}
20 June 1998 \\
\end{flushleft}

\newpage

It is well known that (2+1)-dimensional gauge theories develop unusual 
features, due to the possibility of inclusion of the Chern-Simons (CS) term in 
the action [1]. Especially, due to this CS term, charged matter fields undergo 
unconventional transformation called ``rotational anomaly'' [2] by 
the improved Poincar\'e generators, constructed from the symmetric 
(Belinfante) energy-momentum tensor [3]. Furthermore, it was recently shown 
that the anomalous transformation is uniquely identified as the anomalous 
spin of the matter fields, which can be fractional, due to the unique meaning 
of the improved generators because they are gauge invariant on the constraints
surface and obey the classical Poincar\'e algebra although this is not the 
case for the canonical (Noether) ones [4]. This fact is 
consistent with the fact that the rotational group in (2+1)-dimensions is 
$SO(2)$ which allows only continuous spectrum for the angular momentum. On
 the other hand, without doubt, the anomalous spin is nothing but the 
result of Aharonov-Bohm (AB) effect [5], since the CS term (in the Coulomb 
gauge) makes the charged matter fields carry the point magnetic vortices 
together with their electric charges which is the configuration of the AB 
effect. [This interpretation is still valid with the Maxwell term because  
only the asymptotic behavior is involved and this is controled by the CS 
term [6].] So it has long been believed that this unusual rotational behavior 
is just because one has included the CS term which introduces the AB effect to
the theory.

However, remarkably in a series of paper which are not well-known, Hagen 
showed that even the conventional QED in (2+1)-dimensions, (QED$_3$), also 
produces the rotational anomaly for the charged matter fields [7]. But there 
have been no further studies about the origin of this unexpected result. In 
this Letter, I investigate the physical and mathematical origin of this 
anomaly and relate this anomaly to the Aharonov-Casher (AC) effect [8] and/or 
the AB effect which is not manifest in the Lagrangian. Furthermore, I show 
that there is new anomalies in the time translation generated by Hamiltonian 
for the charged matter fields. Due to the uniqueness of the improved 
Poincar\'e generators, it is shown that this system also develops the 
fractional eigenvalue for the angular-momentum operator. This system shows 
also the attraction (repulsion) for the parallel (anti-parallel) magnetic 
dipole moments contrast to the usual (3+1)-dimensions case. It is shown that 
this system provides a new pairing mechanism for the high-$T_c$ 
superconductivity on the plane.

The model to be studied is QED$_3$ with the massive relativistic complex 
scalars
\begin{eqnarray}
{\cal L} =-\frac{1}{4} F_{\mu \nu} F^{\mu \nu} 
 +(D_{\mu} \phi )^* (D^{\mu} \phi) -m^2 \phi^* \phi,
\end{eqnarray}
where $g_{\mu \nu}$ =diag (1, -1, -1), $F_{\mu \nu}=\partial _{\mu} 
A_{\nu} -\partial _{\nu} A_{\mu}$ and $D_{\mu}=\partial_{\mu} +i q A_{\mu}$. 
( $q$ is the electric charge of the elementary quanta of $\phi$.) The 
equations of motion which follow from (1) are
\begin{eqnarray}
\partial_{\mu} F^{\mu \nu} = J^{\nu}, \\
D_{\mu} D^{\mu} \phi +m^2 \phi =0
\end{eqnarray}
and complex conjugate  of (3). ($J^{\nu}$ is the conserved current $J^{\nu}=i 
q [{\phi^*} D^{\nu} \phi - (D^{\nu} \phi)^* \phi ]$.) For the later purpose, 
let me first consider the (formal) solution of $A^{\mu}$ in (2). In order to 
make the gauge field configurations be rotationally symmetric, I choose the 
Coulomb gauge
\begin{eqnarray}
{\bf \nabla \cdot A }=0.
\end{eqnarray}
Then, from the temporal component of (2), ${\bf \nabla \cdot E }=J^0~ ( 
{\bf E} =-{\bf \nabla} A^0 -\partial_t {\bf A} )$, one obtains the instaneous 
Coulomb potential and it's
multipole expansion around ${\bf x}$
\begin{eqnarray}
A^0({\bf x}, t) &=&\int d^2 {\bf x'}~ G({\bf x}-{\bf x}')~ J^0 ({\bf x}', t) 
\nonumber \\
&=&-\frac{1}{2 \pi}~ Q ~ln |{\bf x}| 
+\sum^{\infty}_{n=2} A^{0}_{(n)} ({\bf x}, t),
\end{eqnarray}
where $G({\bf x}) = -(1/2 \pi)~ ln |{\bf x}|$ is the two spatial-dimensions
 Green's function for the Laplacian, ${\bf \nabla }^2 G ({\bf x}) =-\delta^2(
{\bf x})$ and $Q$ is the total electric charge $Q=\int d^2 {\bf x} J^0 (x)$. 
[There is a constant ambiguity in the definition of $G({\bf x})$, but this 
does not change the main issue of this Letter except in one minor point. This 
will be discussed later.] The direct Taylor expansion of $G({\bf x},t)$ gives 
the multipole expansion of the second line and 
$\sum^{\infty}_{n=2} A^{0}_{(n)} $ represent the (electric) dipole and 
higher-order moments terms and they behave as ${\cal O} (|{\bf x}|^{-1})$ at 
large $|{\bf x}|$. From the spatial components of (2), 
$ \epsilon_{ij} \partial_j B -\partial_t E^i =J^i~ (B={\bf \nabla } \times 
{\bf A} =\epsilon_{ij} \partial_i A^j)$ one obtains the retarded vector 
potential and it's multipole expansion
\begin{eqnarray}
A^i({\bf x}, t) &=&\int d^2 {\bf x'}~ G({\bf x}-{\bf x}') ~{[J^i_T ({\bf x}', 
t')]}_{ret}
\nonumber \\
&=&\frac{1}{2 \pi} \frac{\bf x}{|{\bf x}|^2} \cdot \int d^2 {\bf x'} {\bf x}' 
{[{J}^i_T ({\bf x}', t')]}_{ret}  + \sum^{\infty}_{n=3} A^{i}_{(n)} ({\bf x}, 
t),
\end{eqnarray}
where ${\bf J}_{T}$ is the transverse current $J^i_T \equiv J^i
-\partial_i \partial_t A^0 =-\epsilon_{ij} \partial_j {\bf \nabla  \times} 
\int d^2 {\bf x}' (1/2 \pi) ln |{\bf x}-{\bf x'}| {\bf J}(x')$, $ 
{\bf \nabla \cdot J}_{T} =0$ and $[f({\bf x}',t')]_{ret} =f({\bf x}', 
t-|{\bf x}-{\bf x'}| )$. $\sum^{\infty}_{n=3} A^{i}_{(n)}$ represents the 
expansion of quadrupole and higher-order moments terms and behaves as $
{\cal O} (|{\bf x}|^{-2})$ at large $|{\bf x}|$. [The monopole term was 
neglected as usual by 
considering a well localized $[{\bf J}_T]_{ret}$ [9]] Now using the usual 
mathematical identity [8] $\int d^2 {\bf x'} ({x^i}' {[J^j_T (x')]}_{ret} 
+{x^j}' {[J^i_T(x')]_{ret}} )=0$, (6) becomes
\begin{eqnarray}
A^i ({\bf x}, t)=-\frac{1}{2 \pi} \frac{\epsilon_{ij} x^j }{ |{\bf x}|^2} 
~[\mu ](t;{\bf x}) + \sum^{\infty}_{n=3} A^{i}_{(n)} ({\bf x}, t),
\end{eqnarray}
where $[\mu] =(1/2) \int d^2 {\bf x'} {x}' \times [{\bf J}_{T} ({ x}')]_{ret}$
 is the magnetic moment of the system induced by the retarded current $ 
[{\bf J}_{T} ({x}')]_{ret} $ . Note that $[\mu ]$ has the implicit 
${\bf x}$-dependence through the retardation in $[{\bf J}_{T}(x')]_{ret}$ as 
well as the
$t$-dependence. [These solutions (5)-(7) are only formal ones because one does 
not know the exact solution for $J^{\mu}$ in the classical as well as quantum 
electrodynamics.] With these formal solutions, one gets $B ,~ { E}^i$ as
\begin{eqnarray}
&&B(x)= \frac{1}{2 \pi} \nabla \cdot (\nabla ln|{\bf x}|~ [\mu] )
+\sum^{\infty}_{n=3} \nabla \times {\bf A}_{(n)} , \nonumber \\
&&E^{i}(x) =\frac{1}{2 \pi} \frac{x^i}{|{\bf x}|^2} Q +
\frac{1}{2 \pi} \frac{\epsilon_{ij}x^j}{|{\bf x}|^2}~ [\dot{\mu}] 
+\sum^{\infty}_{n=3}[-\partial_i A^{0}_{(n-1)}-\partial_t A^i_{(n)}],
\end{eqnarray}  
where over-dot represents the time derivative and higher-order 
moments terms of $B$ and $E^i$ behave as ${\cal O}(|{\bf x}|^{-3})$ and 
${\cal O}(|{\bf x}|^{-2})$ at large ${\bf x}$, respectively.

Now, let me consider the canonical quantization with the usual commutation 
relations [2, 10]
\begin{eqnarray}
&&[ \phi({\bf x}, t), \pi({\bf x}',t) ] 
=[ \phi^{\dagger}({\bf x}, t), \pi ^{\dagger}({\bf x}',t) ]=i \delta^2 
({\bf x}-{\bf x}'), \nonumber \\
&&[A^i({\bf x}, t), E^j ({\bf x}', t) ]=-i \delta ^{T}_{ij} 
({\bf x}-{\bf x}'), \\
&&\mbox{others~ vanish}, \nonumber 
\end{eqnarray}
with $\pi=(\partial_t -i q A_0) \phi^{\dagger},~\pi^{\dagger}=\phi
(\stackrel{\leftarrow}{\partial_t} 
+i q A_0),~\delta_{ij}^T ({\bf x})=(\delta_{ij} -\partial_i \partial_j 
{\nabla }^{-2} )\delta ({\bf x})$. [From now on, all quantum commutators are 
assumed to be the equal-time commutators.] To examine the Poincar\'e 
transformation and the covariance issues, let me consider the (manifestly) 
gauge invariant 
and symmetric energy-momentum tensor
\begin{eqnarray}
T^{\mu \nu}_{s} =(D^{\mu} \phi)^* (D^{\nu} \phi) +(D^{\nu} \phi)^* (D^{\mu} 
\phi) + F^{\mu \rho} {F_{\rho}}^{\nu}-g^{\mu \nu} \left[
 (D_{\mu} \phi )^* (D^{\mu} \phi) -m^2 \phi^* \phi
-\frac{1}{4} F_{\sigma \rho} F^{\sigma \rho} \right].
\end{eqnarray} 
From this, one can obtains the improved Poincar\'e generators, which are the 
gauge invariant constants of motion [3]
\begin{eqnarray}
&&P^0_s = \int d^2 {\bf x} \left[ \pi \pi^{\dagger} +({\bf D} \phi) \cdot (
{\bf D} \phi)^{\dagger}+ m^2 \phi \phi^{\dagger}+\frac{1}{2} ({\bf E}^2 + B^2)
 \right], \nonumber \\
&&P^i_s =\int d^2 {\bf x} \left[ \pi D^i \phi +(D^i \phi)^{\dagger} \pi^{
\dagger}  -\epsilon_{ij} E_j B \right], \nonumber \\
&&M^{12}_s =\int d^2 {\bf x} ~\epsilon^{ij} x^i \left[ \pi D^j \phi + 
(D^j \phi)^{\dagger}   \pi^{\dagger} -\epsilon_{jk} E_k B \right], 
\nonumber \\
&&M^{0i}_s =t P^i_s -\int d^2 {\bf x} ~x^i \left[ \pi \pi^{\dagger} 
+({\bf D} \phi) \cdot ({\bf D} \phi)^{\dagger}+m^2 \phi \phi^{\dagger} 
+\frac{1}{2} 
({\bf E} ^2 + B^2) \right],
\end{eqnarray}
by $P^{\mu}_s=\int d^2 {\bf x} T^{0 \mu}_s,~ 
M^{\mu \nu}_s =\int d^2 {\bf x} (x^{\mu} T^{0 \nu}_s-x^{\nu} T^{0 \mu}_s )$. 
These are the well-known forms and produce the usual (first-order) Heisenberg 
equations and satisfy the Poincar\'e algebra in (3+1)-dimensions with the 
symmetric ordering. But as one shall see, this is not a trivial matter in 
(2+1)-dimensions due to the non-trivial surface term contributions. First of 
all, if one consider the spatial 
translation of the matter fields generated by $P^i_s$, it is easy to verify 
that one obtains the usual result 
$i [ P^i_s,\phi(x)] =\partial^i \phi (x)$, where I have dropped the surface 
term $\int
d^2 {\bf x'} \{ ~\partial ' _i ~[~({\bf \nabla}' ln |{\bf x}-{\bf x}'|) 
{\bf \cdot A}(x')~] -{\bf 
\nabla ' \cdot}[~({\bf \nabla}' ln |{\bf x}-{\bf x}'|) A^i(x')~ ] - {\bf 
\nabla '\cdot}[~(\partial_i ' ln |{\bf x}-{\bf x}'|) {\bf A}(x') ~]~\}$ 
because it vanishes as $|{\bf x'}|^{-1},~|{\bf x'}| \rightarrow \infty$ using 
the asymptotic behavior of ${\bf A}$ in (7). Next, if one consider the time 
translation generated by $P^0_s$, one gets
\begin{eqnarray}
i[P^0_s,\phi(x)] =\dot{ \phi}- i\frac{q^2}{4 \pi} ln({\bf 0})~ \phi 
-i\frac{q}{4 \pi} lnR|_{R \rightarrow \infty} [Q, \phi]_{+},
\end{eqnarray}
where $[A, B]_{+}=AB+BA$. The second term is a quantum correction arising from
reordering at the same point [11] and is absent if one follows the old 
formulation of Pauli and Weisskopf without choosing the gauge [10]. Moreover 
this term appears only
in the bosonic theory because of $(A^0)^2$ term in $P^0$. 
The third term, which exists even in the classical level, is a surface 
contribution 
from $\int d^2 {\bf x'}~ {\bf \nabla ' \cdot (E }(x') ln |{\bf x}-{\bf x}' | )
=  Q~ln R |_{R \rightarrow \infty} $ using the asymptotic behavior of 
${\bf E}$ in (8). This is a kind of the infrared divergence which is 
independent on the mass of the matter field in (2+1)-dimensions. Here, it is 
noted that although (12) contains two unconventional terms, the second-order 
quantum equations of motion are the same 
form as the classical ones (3). This implies that the commutation relations 
(9) are a consistent choice. On the other hand, it is easy to see that the 
modified generator $\widetilde{P^0_s}=\int d^2 x \widetilde{T^{00}_s}$ with
\begin{eqnarray}
\widetilde{T^{00}_s}(x) ={T^{00}_s(x)} -\left[ \frac{q}{4 \pi} ln({\bf 0}) 
J^0(x) +{\bf \nabla \cdot } \left( \frac{Q^2}{2 (2 \pi)^2 } \frac{\bf x}
{|{\bf x}|^2} ln |{\bf x}| \right) \right]
\end{eqnarray}
gives the usual transformation $i[\widetilde{P^0_s},\phi(x) ]=\dot {\phi}$ 
but this is not allowed: The first term in the bracket (13) cancels the 
reordering effects in (12) and (14), but it breaks the Poincar\'e algebra 
involving 
$\widetilde {M}^{0i}_s=\int d^2 {\bf x}~ (t T^{0i}_s-x^i \widetilde{T^{00}_s} 
)$. Moreover, although the second term
does not have this problem, $\widetilde{T^{\mu \nu}_s}$ does not transform as 
the second-rank tensor under the boost operator, i.e., 
$i [ \widetilde{T^{00}}, M^{0i}] =(t \partial^i -x^i \partial_t) 
\widetilde{T^{00}} +2 T^{0i} - t (Q^2/8 \pi^2) \partial^i \nabla \cdot 
(\frac{\bf x}{|{\bf x}|^2} ln |{\bf x}|)$. These two facts are connected with 
the non-covariant forms of the vacuum expectation values of the the added 
terms in (13) [12]. Especially for the reordering term, it can be absorbed 
into the gauge transformation $\phi \rightarrow \phi e^{i \alpha t} \equiv 
\phi', ~ A_0 \rightarrow A_0 -\frac{\alpha}{q} \equiv A_0 '$ with 
$\alpha=-(q^2/4 \pi) ln({\bf 0})$, but it is questionable that the fields 
$\phi'$, which has no reordering effect, is meaningful one because it 
oscillates with infinite frequency. So, the two modification terms of (13) 
are not  
physical ones in these senses. 

The Lorentz boost generated by $M^{0i}_s$ becomes
\begin{eqnarray}
i[M^{0i}_s, \phi(x) ] &=&(t \partial^i-x^i \partial_t ) \phi(x) 
-i\frac{q}{2 \pi} \int d^2 {\bf x'}~ ln |{\bf x }-{\bf x}'| \left[ E^i (x') 
+(x'^i -x^i) J^0(x'), \phi (x) \right]_{+} \nonumber \\
 &+&\frac{i q ^2}{4 \pi} x^i ln ({\bf 0}) ~\phi(x).
\end{eqnarray}
Here, I have neglected the surface contribution which comes from
\begin{eqnarray}
\int d^2 {\bf x'}~ {\bf \nabla ' \cdot} (x'^i {\bf E} (x') ln |{\bf x}
-{\bf x}'| ) =- Q ~\oint _{S^1_{R\rightarrow \infty}} d \theta {\bf\hat{r}}^i
 R ln R
\end{eqnarray}
[ The integration is evaluated on a circle with radius $R$, polar angle 
$\theta$, and their corresponding orthogonal unit vectors ${\bf \hat{r}} 
=(cos \theta, sin \theta), {\bf \hat{\theta}}$. ] since it vanishes for 
any index $i$ because of symmetry. However, note that the naive power 
counting gives $R ln R |_{R \rightarrow \infty}$ divergence. So, in this case 
there is no surface contribution contrast to the naive expectation arising 
from the result (12). Returning to the result (14), the second term is the 
usual Lorentz transformation term for the matter field to accommodate with 
the Lorentz transformation of the gauge field ${\bf A}$ in the Coulomb gauge 
[13]. The third term is the reordering effect in the same origin with that of 
(12).

Finally, the rotation generated by $M^{12}_s$ becomes
\begin{eqnarray}
i[M^{12}_s, \phi(x)] &=&{\bf x \times \nabla} \phi +i\frac{q}{2 \pi} 
\int d^2
{\bf  x'}~ {\bf \nabla ' \times }\left[~ {\bf A}(x')~ 
( {\bf x' \cdot \nabla '} 
ln|{\bf x}-{\bf x'}|) ~\right] \phi(x)  \nonumber \\
&=&{\bf x \times \nabla} \phi -i\frac {q}{2 \pi} [\mu](t; |{\bf x}|)|_{
{\bf x}\rightarrow \infty }~ \phi
\end{eqnarray}
using the asymptotic form of ${\bf A}$ in (7) and the commutation relation 
$[[\mu], \phi ] =0$ which is implied by $[ A^i, \phi]=0$ in (9). Moreover,
 because of the fact that $[\mu](t;{\bf x}) |_{ |{\bf x}| \rightarrow \infty} 
=(1/2) 
\int d^2 {\bf x'} ~{\bf x' \times J_{T}}({\bf x'}, t-|{\bf x}|) |_{ |{\bf x}| 
\rightarrow \infty} \equiv \mu (t-\infty)$ and the anomalous term in (16) 
depends on the initial 
magnetic moment at the temporal infinity $t'=-\infty$, i.e., 
$\mu|_{t'=-\infty} 
\equiv \mu_{-\infty}$ for any finite time $t$. Here, there is no reordering 
effect in the same way as $P^i_s$ but the surface contributions which were 
neglected in $[ P^i_s, \phi]$ can not be neglected  because of the existence 
of one more spatial coordinate in the integrand of 
$M^{12}_s$. This unconventional term which is fixed by the initial condition 
of the system at $t'=- \infty$, gives the anomalous spin for the matter field.
This anomalous rotation transformation was previously noticed by Hagen [7]
but the origin of this has been unclear. In order to understand the physical 
origin of the anomalous term more clearly, it is helpful to consider the 
surface contribution in the generator itself as
\begin{eqnarray}
M^{12}_s &=&M^{12}_c -\int d^2 {\bf x}~ {\bf \nabla \cdot}[{\bf E}~
({\bf x \times A})] \nonumber \\
&=&M^{12}_{c}  +\oint _{S^{1}_{R\rightarrow \infty}} d \theta R^2 {\bf 
{\hat r} 
\cdot E}~({\bf {\hat r} \times A}) \nonumber \\
&=&M^{12}_{c} +\frac{1}{2 \pi} Q \mu_{-\infty} 
\end{eqnarray}
from which (16) is readily seen to follows: The first term is the canonical 
angular momentum operator $M^{12}_c= \int d^2 {\bf x}~ [ {\bf x \times }
( \pi {\bf \nabla} \phi +({\bf \nabla} \phi)^* \pi^* -E_k {\bf \nabla} A_k ) 
+{\bf E \times A }]
$ which produces the usual angular momentum part and the second term gives the
anomalous contribution of (16). Here, note that only $E_r ={\bf {\hat r} 
\cdot E}$ and 
$A_{\theta}={\bf {\hat r} \times A }$ at the spatial infinity contribute to 
the surface term. But, from the 
asymptotic forms (7) and (8), only the field configurations 
${\bf E}_{R \rightarrow \infty}=(1/2 \pi) Q
 ~{\bf {\hat r}}/R , ~ {\bf A}_{R \rightarrow \infty} =
-(1/2 \pi)\mu _{-\infty} ~ {\bf {\hat \theta}} /R $ are relevant [Since the 
next order 
terms in (7), (8) give the vanishing contribution ${\cal O} (R^{-1})$ to the 
surface term in (17) as 
$R \rightarrow \infty $.] and they are the configurations for the plane 
projection 
of the infinite line charges (charge density $Q$ per unit length) with a 
magnetic dipole moment ${\bf \mu} =\mu_{-\infty} {\bf {\hat z}}$ in 
(3+1)-dimensions. 
This interpretation implies that the physical origin of the unconventional 
contribution in (16) is the AC effect [7, 14] (when the configuration is 
considered as that of the motion of magnetic dipole $ \mu _{-\infty}$ in the 
electric field ${\bf E}_{R \rightarrow \infty}$) or the AB effect (when the 
configuration is considered as that of motion of charge $Q$ in a magnetic 
field ${\bf \nabla \times A}_{R \rightarrow \infty}$). This fact is not 
manifest in the Lagrangian (1). Furthermore, since there is no reason for 
quantization of ${\mu}_{-\infty}$, the anomalous spin 
$s=-(1/2 \pi)q \mu_{-\infty}$ 
can be fractional. This is the only known example of fractionally-valued 
total angular momentum in the parity-invariant gauge theory in 
(2+1)-dimensions without recourse to the CS term [15]. 

In order to give a unique meaning to the anomalous spin we must verify that 
the improved generators of (11) have more preferred meaning than the 
canonical ones. Usually, these two types of generators are identical on the 
constraint surface when one drops the surface term as in the (3+1)-dimensions 
case. But this is not true in (2+1)-dimensions in general; (17) is an example. 
Actually in this case the differences of these two types of generators, other 
than rotation operator, 
become 
\begin{eqnarray}
P^0_c -P^0_s &=&\int d^2 {\bf x}~ {\bf \nabla \cdot }({\bf E} A^0 )=
-\frac{1}{2 \pi} ln R |_{R \rightarrow \infty} Q^2, \nonumber \\
P^i_c -P^i_s &=&\int d^2 {\bf x}~ {\bf \nabla \cdot }({\bf E} A^i)=0, 
\nonumber \\
M^{0i}_c-M^{0i}_s&=&\int d^2 {\bf x} ~{\bf \nabla \cdot }[ (t A^i - x^i A^0 ) 
{\bf E} ]=0, 
\end{eqnarray} 
where I have used the asymptotic forms of (5)-(8) and the fact that 
geometrical term (15) vanishes. [If one allow the constant ambiguity in the 
Green's function as $G({\bf x}) =-(1/2 \pi) ln |{\bf x}| + const. $, one can 
make $P^0_c =P^0_s$ by choosing the constant as $(1/2 \pi) ln R |_{R
 \rightarrow \infty}$ but this has no effect in the physics, i.e., the 
anomalous contribution in (12). Actually this is the only modification 
produced by the constant ambiguity, which is only a minor point in our 
analysis] If one consider the (3+1)-dimensions, it is easy to verify that 
these surfaces terms all vanish as $P^0_c -P^0_s ={\cal O}(R^{-1})|_{R
\rightarrow \infty}~, P^i_c -P^i_s ={\cal O}(R^{-2})|_{R\rightarrow \infty}, 
M^{ij}_c -M^{ij}_s ={\cal O}(R^{-1})|_{R\rightarrow \infty}$ by power counting
 and $M^{0i}_c -M^{0i}_s=0$ by the geometrical reason 
essentially the same as (15) [16]. So, the fact of inequivalence of the two 
types of the generators is a genuine effect in (2+1)-dimensions. A similar but 
not exactly the same things happen when there is the CS term instead of the 
Maxwell term [4]. Hence, in (2+1)-dimensions it seems that there are at least 
two inequivalent spin contents for the matter fields depending on the type of 
the angular momentum operator: One is integer from the canonical operator 
$M^{12}_c$ and the other is fractional from the improved ones $M^{12}_s$. 
However, as can be verified by tedious calculations, it is not the canonical 
ones but the improved generators with the orderings in (11), which obey the 
quantum Poincar\'e algebra: In the canonical case, the commutator of Lorentz 
boost becomes $[M^{0i}_c, M^{0j}_c] =-i \epsilon_{ij} (M^{12}_c +({1}/{2 \pi})
 \mu_{-\infty} Q)$ and the anomalous term on the right-hand side can not be 
removed without modifying other generators. 
[ This can be also easily verified by noting that the gauge invariant fields 
combinations ${\bf E}, B, \pi \pi^{\dagger}, ({\bf D} \phi )\cdot 
({\bf D} \phi)^{\dagger}$, 
\ldots etc., only by which the Poincar\'e generators can be expressed, have 
no anomalous transformations and no gauge dependent Lorentz transformation 
term
contrast to the charge bearing fields $\phi, \phi^{\dagger}$ and the gauge 
field ${\bf A}$. So the calculation with the gauge invariant fields should be 
the same as in (3+1)-dimensions case where the generators satisfy the usual 
Poincar\'e algebra.] From this fact the canonical generators can be discarded 
as the unphysical ones. Moreover, from the fact that the anomalous term in 
$M^{12}_s$ can not be discarded by a redefinition of generators [2], the 
improved generators have a unique meaning consistently with Einstein's theory 
of gravity [17]. 
Hence the anomalous spin which comes only from $M^{12}_s$, has a unique 
meaning also. However, this uniqueness of anomalous spin does not imply the 
anomalous statistics, which has only artificial meaning similar to the CS 
gauge theory [4].

Finally, I discuss one more unusual phenomenon of our model and it's 
implication to the high-$T_c$ superconductivity. If one imagine the
 non-relativistic identical point-particles (these need not be restricted to 
the bosons) with the same magnetic moment 
$\mu$ [I have omitted the retardation symbol [~ ] because there is no 
retardation effect 
in this non-relativistic case], it gives an attractive $\delta$-potential in 
the dipole approximation 
\begin{eqnarray}
U({\bf x}) \cong - \mu B \cong - \mu ^2 \delta({\bf x}) .
\end{eqnarray}
The oppositely oriented (i.e., opposite sign) magnetic dipoles are repulsive 
[18]. 
This is sharply contrast to the phenomena in (3+1)-dimensions: In 
(3+1)-dimensions, the corresponding potential for two identical dipoles is 
$U({\bf x})
\cong -{\bf \mu \cdot B} \cong {\bf \mu \cdot \mu }/(4 \pi |{\bf x}|^3) 
-(2 \pi/ 3) {\bf \mu \cdot \mu} \delta^3 ({\bf x})$ and so a collapse of 
the dipoles is avoided from the existence of infinite barrier represented by 
the first term; But there is no infinite barrier in (2+1)-dimensions case 
(19) such that the collapse is inevitable. Moreover, it is known that the quantum mechanics allows a stable bound 
state with the binding energy $\Delta =\epsilon/(e^{m/\mu^2}-1)$ for the 
two-body system with the potential (19) [19]. [The stable $N$-body ground state can be also constructed
and is correspond to a 
Hartree-Fock-type approximation.] Here, $\epsilon$ is the energy cut-off. 
Hence, this system provides a new pairing mechanism for the planer 
systems, different from the Cooper-pairing mechanism [20]: In this new 
mechanism, the `` ferromagnetic order '' is favored as in the recent {\it 
magnon exchange pairing mechanism} for the high-$T_c$ $CuO$ superconductors 
[21].
Actually, it is easy to compute a rough estimate of the necessary magnetic 
moment $\mu \cong ln 2 ~\mu_B$ ($\mu_B =e/2 m_e$ is the Bohr-magneton) in 
order to give a critical temperature $T_c \cong \Delta/ k_B \cong 200~ K$ 
($k_B$ is the Boltzmann constant) between normal (non-binding) and 
superconducting (binding) phases. [Here, $m$ and $\epsilon$ are chosen as the 
electron mass $m_e$ and it's rest energy, respectively] The time-reversal and parity 
symmetries of the system are consistent with a recent experimental result 
[22].     

In summary, I have shown that QED$_3$ reveals several unusual phenomena due to
 the surface term contributions. Now, let me address several comments. First, 
about the AC effect which was identified from the surface term in $M^{12}_s$, 
this can be also expected by noting that the matter field $\phi$ in the 
Coulomb gauge carries the radial electric field ${\bf {\cal E} }({\bf x'})=-
(q/2 \pi) ({\bf x}-{\bf x}')/({\bf x}-{\bf x}')^2$ because of $[{\bf E}
({\bf x}', t), \phi({\bf x}, t) ]={\bf {\cal E}}({\bf x}') \phi({\bf x}, t)$, 
which is one needed constituent for the AC effect. And as one more necessary 
consitutent, the magnetic moment is provided by the conserved  current at 
$t'=-\infty$. One notable thing is that in the fermion case, the intrinsic 
magnetic moment of the fermion does not contribute additionally to the AB/AC 
effect through the surface term in (16) or (17): This is a completely 
analogous situation to the AC effect in CS gauge theory [2]. Moreover, if one 
includes 
the CS term in the Lagrangian (1), the AB effect, which comes from the CS 
term, 
appears rather than the AB/ AC effect which comes from the Maxwell term: In 
this case the lower-derivative CS term will dominate the higher-derivative
 Maxwell term asymptotically [9, 23]. As for the superconductivity, it would 
be interesting to consider the finite temperature field theory to get a fully 
field theoretical description of the phenomena.
As a final comment, although I have considered only the Coulomb gauge, the 
anomalous surface contributions, i.e., the anomalous field transformation and 
the 
anomalous spin are gauge independent: This is  because  the surface terms are 
gauge invariant for the rapidly 
decreasing gauge transformation function $\Lambda$ under the gauge 
transformation $A_{\mu} \rightarrow A_{\mu} +\partial_{\mu}\Lambda, ~ 
\phi \rightarrow exp[-iq \Lambda ] \phi $ and the commutator relation $[Q, 
\phi(x) ] =-q \phi(x)$ is gauge independent [4].\\

I would like to thank Prof. S. Deser, C. R. Hagen, R. Jackiw, J. D. Jackson, 
I. Kogan for valuable conversations and informations and Prof. Y.-J. Park for 
warm hospitality. I also acknowledge the financial support of Korea Research 
Foundation made in the program year 1997. The present work is supported also 
in part by funds provided by the U.S. Department of Energy (D.O.E) under 
cooperative research agreement No. DF-FC02-94ER40818.

\newpage
\begin{center}
{\large \bf References}
\end{center}
\begin{description}

\item{[1]} S. Deser, R. Jackiw, and S. Templeton, Phys. Rev. Lett. 
{\bf 48}, 975 (1982); Ann. Phys. (N.Y.) {\bf 140}, 372 (1982).

\item{[2]} C. R. Hagen, Ann. Phys. (N.Y.) {\bf 157}, 342 (1984);
Phys. Rev. {\bf D31}, 2135 (1985); A. M. Polyakov, Mod. Phys. Lett. {\bf A3},
 325 (1988); G. W. Semenoff, Phys. Rev. Lett. {\bf 61}, 517 (1988).

\item{[3]} F. J. Belinfante, Physica (Utrechet) {\bf 7}, 449 (1940).

\item{[4]} M.-I. Park and Y.-J. Park, `` New Gauge Invariant Formulation of 
the Chern-Simons Gauge theory '', MIT-CTP-2702, hep-th/9803208 (1998).

\item{[5]} W. Ehrenberg and R. E. Siday, Proc. Phys. Soc. London, Sec. 
{\bf B62}, 8 (1949); Y. Aharonov and D. Bohm, Phys. Rev. {\bf 115}, 485 (1959).

\item{[6]} R. Jackiw, Ann. Phys. (N.Y.) {\bf 201}, 83 (1990).

\item{[7]} C. R. Hagen, Phys. Rev. Lett. {\bf 58}, 1074 (1987); Phys. Rev. 
{\bf D36}, 3773 (1987).

\item{[8]} Y. Aharonov and A. Casher, Phys. Rev. Lett. {\bf 53}, 319 (1984).

\item{[9]} J. D. Jackson, {\it Classical Electrodynamics } (John Wiley and 
Sons, Inc., New York, 1962).

\item{[10]} W. Pauli and V. Weisskopf, Helv. Phys. Acta, {\bf 7}, 709 (1934); 
J. D. Bjorken and S. D. Drell, {\it Relativistic Quantum Fields } 
(McGraw-Hill, Inc., New York, 1965).

\item{[11]} R. Jackiw and S. Y. Pi, Phys. Rev. {\bf D42}, 3500 (1990);
M.-I. Park and Y.-J. Park, {\it ibid.}, {\bf D50}, 7584 (1994).

\item{[12]} D. G. Boulware and S. Deser, J. Math. Phys. {\bf 8}, 1468 (1967).

\item{[13]} B. Zumino, J. Math. Phys. {\bf 1}, 1 (1960).

\item{[14]} C. R. Hagen, Phys. Rev. Lett. {\bf 64}, 2347 (1990).

\item{[15]} C. R. Hagen, Phys. Rev. Lett. {\bf 68}, 3821 (1992); M. E. 
Carrington and G. Kunstatter, Phys. Rev. {\bf 51}, 1903 (1995).

\item{[16]} Recently, this kind of things in (3+1)-dimensions have been 
questioned by Lavelle and McMullan. But our explicit analysis shows that 
this is not true due to the geometrical reason; see M. Lavelle and 
D. McMullan, Phys. Rep. {\bf 279}, 1 (1997).

\item{[17]} Motivated by the modified gravity theory as suggested by Callan
 $et~ al.$, one can consider the new improved generators, 
$\bar{P}^{\mu} =\int d^2 
{\bf x} ~\bar{T}^{0 \mu},~ \bar{M}^{\mu \nu} =\int d^2 {\bf x}~( x^{\mu} 
\bar{T}^{0 \nu}-
x^{\nu} \bar{T}^{0 \mu})$ with $\bar{T}^{\mu \nu}=T^{\mu \nu}+ c
(g ^{\mu \nu} \Box-\partial^{\mu} \partial^{\nu}  )f(\phi^* \phi)$. 
Here, c is constant and $f(\phi ^* \phi)$ is some function of $\phi^* \phi$. 
But it is easy 
to verify that $\bar{P}^{\mu}=P^{\mu}, ~\bar{M}^{\mu \nu}=M^{\mu \nu}$, with 
reasonable assumptions a) no angle dependence of $f(\phi ^* \phi)$  and
 b) $f(\phi^* \phi) \sim |{\bf x}|^{-\delta}~ (\delta > 0)$ as $|{\bf x}| 
\rightarrow \infty$. So, there is no change in the conclusion of this Letter; 
see C. G. Callan, S. Colman, and R. Jackiw,
Ann. Phys. (N.Y.) {\bf 59}, 42 (1982).

\item{[18]} A similar behavior was studied by Stern but exact identification 
remains to be clarified; see J. Stern, Phys. Lett. {\bf B 265}, 119 (1991).

\item{[19]} C. Thorn, Phys. Rev. {\bf D19}, 639 (1979); K. Huang, 
{\it Quarks, Leptons and Gauge Fields} (World Scientific, Singapore, 1982); 
R. Jackiw, in { M. A. M. B\'eg Memorial volume} (World Scientific, Singapore, 1991).

\item{[20]} L. N. Cooper, Phys. Rev. {\bf 104}, 1189 (1956).

\item{[21]} G. Chen and W. A. Goddard III, Science {\bf 239}, 899 (1988).

\item{[22]} S. Spielman {\it et al.,} Phys. Rev. Lett. {\bf 65}, 123 (1990).

\item{[23]} K. Shizuya and H. Tamura, Phys. Lett. {\bf B 252}, 412 (1990).
\end{description}
\end{document}